\documentclass[a4paper,12pt]{article}

\usepackage{graphics}

\begin{document}

%\begin{flushright}
%{\normalsize IRB-TH-4/99 \\
%July, 1999.}
%\end{flushright}
\vspace{1cm}
\thispagestyle{empty}
\title{
\vspace{-2cm}
\begin{flushright}
{\normalsize IRB-TH-4/99 \\ 
\vspace{-0.5cm}
July, 1999.}
\end{flushright}
%\vspace{2 cm}
\bf Enhancement of preasymptotic effects in inclusive beauty decays}

\author{B. Guberina\thanks{guberina@thphys.irb.hr},
 B. Meli\'{c}\thanks{melic@thphys.irb.hr},
 H. \v Stefan\v ci\'c\thanks{shrvoje@thphys.irb.hr}
}

\vspace{2cm}
\date{
\centering
Theoretical Physics Division, Rudjer Bo\v{s}kovi\'{c} Institute, \\
   P.O.Box 1016, HR-10001 Zagreb, Croatia}

%\institute{
%  Theoretical Physics Division, Ru\dj er Bo\v{s}kovi\'{c} Institute,
%   P. O. Box 1016, HR-10001 Zagreb, Croatia}

\maketitle

{\abstract We extend Voloshin's recent analysis of charmed and beauty
hyperon decays based on $SU(3)$ symmetry and heavy quark effective theory,
by introducing a rather moderate model-dependence, in order to obtain more
predictive power, e.g. the values of lifetimes of the
$(\Lambda_{b},\Xi_{b})$ hyperon triplet and the lifetime of $\Omega_{b}$. In
this way we obtain an improvement of the ratio
$\tau(\Lambda_{b})/\tau(B_{d}^{0}) \sim 0.9$ and the hierarchy of lifetimes
$\tau(\Lambda_{b}) \simeq \tau(\Xi_{b}^{0}) < \tau(\Xi_{b}^{-}) <
\tau(\Omega_{b})$ with lifetimes of $\Xi_{b}^{-}$ and $\Omega_{b}$
exceeding the lifetime of $\Lambda_{b}$ by $22 \%$ and $35 \%$,
respectively.}

\vspace{2cm}


\newpage

\indent
Weak decays of beauty mesons and baryons are believed to be a nice
playground where a variety of phenomena should be well described and
understood in the framework of the operator product expansion (OPE) and
heavy quark effective theory (HQET) \cite{Neu,BlokShif}.
The essential underlying idea in both theories
is the expansion in inverse powers of heavy-quark mass -- the mass of the
beauty quark, $m_{b} \sim O(5 \, GeV)$, is considered to be heavy compared
with
the typical hadron scale of $0.5-1 \,GeV$. This is to be compared with the
case of charmed mesons and baryons, where the mass of the charmed quark,
$m_{c} \sim 1.3 \,GeV$, is hardly an ideal expansion parameter.

The rate of the beauty-hadron decay is given by
\begin{eqnarray}
\label{eq:master}
 \Gamma (H_{b} \rightarrow f) &=& \frac{G_{\rm F}^2 m_b^5}{192 \pi^3} |V|^2
\frac{1}{2 M_{H_{b}}} \{ c_3^f \langle H_{b}|O_{3}|H_{b}\rangle +
c_5^f \frac{ \langle H_{b}|O_{5}|H_{b}\rangle }{m_b^2}  \nonumber \\
 &+& \sum_i c_6^f \frac{ \langle H_{b}|O_{6}^{i} |H_{b}\rangle }{m_b^3} 
+ O(1/m_b^4) + ...\}\, ,
\end{eqnarray}

where $c_{j}^{f}$ are the Wilson coefficients and
\begin{equation}
\label{eq:matrel}
\frac{1}{m_{b}^{D-3}} \langle H_{b}|O_{D}|H_{b}\rangle 
\end{equation}
are matrix elements of the D-dimensional operators which appear in the OPE
multiplied by the appropriate power of inverse quark mass.
The sum in (\ref{eq:master}) starts with $D=3$, i.e. with
$O_{3}=\overline{b} b$ giving
\begin{equation}
\label{eq:op3}
\frac{1}{2 M_{H_{b}}} \langle H_{b}|O_{3}|H_{b}\rangle = 1 + O(1/m_{b}^{2}) \,.
\end{equation}
Clearly, in the asymptotic limit $m_{b} \rightarrow \infty$, one recovers
the parton model result -- as long as $m_{b}$ is large enough, one expects
all corrections to stay moderate. Furthermore, it is obvious from
(\ref{eq:op3}) that there are no  $1/m_{b}$ corrections -- a
consequence of the nonexistence of independent operators of dimension four.

The experimental situation is as follows: the lifetimes of beauty hadrons
follow the simple theoretical $m_{b} \rightarrow \infty$ 
prediction within $5-10 \%$:
\begin{equation}
\label{eq:lifetimes}
\tau(B^{+})=\tau(B_{d}^{0})=\tau(B_{s}^{0})=\tau(\Lambda_{b}) \, ,
\end{equation}
except for the lifetime of $\Lambda_{b}$, which appears to be by $15-25 \%$ 
smaller than predicted in (\ref{eq:lifetimes}). More precisely \cite{PDG},
\begin{equation}
\label{eq:ratioexp1}
\frac{\tau(B^{+})}{\tau(B_{d}^{0})} = 1.07 \pm 0.03 \, ,
\end{equation}
\begin{equation}
\label{eq:ratioexp2}
\frac{\tau(\Lambda_{b})}{\tau(B_{d}^{0})} = 0.81 \pm 0.05 \, .
\end{equation}

The lifetimes of $b$ hadrons are 
\begin{eqnarray}
\label{eq:totals}
\tau(B_{d}^{0}) = (1.54 \pm 0.03) \,ps \, , \nonumber \\
\tau(B^{+}) = (1.65 \pm 0.03) \, ps \, , \nonumber \\
\tau(\Xi_{b} \; mixture) = (1.39 
%\stackrel{\displaystyle +0.34}{-0.28}
\begin{array}{c}
\vspace{-0.2cm} 
\scriptstyle +0.34 
\vspace{-0.3cm}
\\
\scriptstyle -0.28
\vspace{0.2cm}
\end{array}
) \, ps \nonumber \\
\tau(\Lambda_{b}) = (1.24 \pm 0.08) \, ps \, .
\end{eqnarray} 

Theoretical estimates \cite{ShifB} predict the ratio
(\ref{eq:ratioexp1}) to be $1+0.05(f_{B}/200 MeV)^{2}$,  
%
%\begin{equation}
%\label{eq:Shif}
%1+0.05(\frac{f_{B}}{200 MeV})^{2}
%\end{equation}
%
in accordance with experiment, but the ratio (\ref{eq:ratioexp2}) is
predicted to be in the range
\begin{equation}
\label{eq:ratiotheor}
\frac{\tau(\Lambda_{b})}{\tau(B_{d}^{0})} \sim 0.95-0.98 \, ,
\end{equation}
which seems to be an overestimate.

It appears, however, that the ratio
$\tau(\Lambda_{b})/\tau(B_{d}^{0})$ is not easy to lower down to the
experimental value, the reason being that the $1/m_{b}$ expansion
converges rapidly. For example, keeping only
operators with $D=3$ and $D=5$, one obtains $0.98$ for the ratio
(\ref{eq:ratioexp2}).
Thus it seems difficult to accommodate this ratio with the same mass $m_{b}$ 
entering the
decay rates of both $\Lambda_{b}$ and $B_{d}^{0}$. In fact, strangely
enough, putting the physical hadron masses instead of $m_{b}$
would give
$\tau(\Lambda_{b})/\tau(B_{d}^{0}) = 0.73$,  up to the $O(1/m_{b}^{2})$,
%
%\begin{equation}
%\label{eq:Alt}
%\frac{\tau(\Lambda_{b})}{\tau(B_{d}^{0})} = 0.73
%\end{equation}
%
in good  agreement with experiment. However, this nice Ansatz, proposed in
\cite{Alt} completely spoils the OPE and contradicts other
OPE predictions confirmed by experiments. 

Therefore, the only hope to
obtain the ratio (\ref{eq:ratioexp2}) in the framework of the OPE and HQET is to
look for the possible larger contributions coming from the operators with
dimension $D=6$ or higher.
These operators are known to play an important role in charmed-meson
decays, in which, owing to the Pauli interference effect \cite{Gub1, Gub2,
Shifman1}, there is a dilation of the lifetime of the $D^{+}$ meson. In 
charmed-baryon decays, their role is even more pronounced:
 they give the dominant
contribution leading to the well-established lifetime hierarchy which was 
successfully predicted prior to experiment \cite{Shifman2, Gub3}.  
Unfortunately, the calculation of the matrix
elements of the operators with dimension $D=6$ requires the use of quark
models and is, therefore, strongly model dependent.

Recently, Voloshin \cite{VolSU3} proposed the way to avoid the use of
phenomenological models. He showed that using $SU(3)$
symmetry and HQET it was possible 
to relate the measured lifetimes of charmed hyperons to
the differences in semileptonic decay rates, the differences in the Cabibbo
suppressed decay rates of charmed hyperons and the splitting of the total
decay rates of $b$ hyperons, without invoking the quark model results for
the matrix elements \cite{Vol, GM}. He confirmed the predicted difference in
the semileptonic decay rates between the $\Xi_{c}$ and $\Lambda_{c}$ by a
factor $2$ to $3$, and the enhancement of the semileptonic branching ratio
for $\Lambda_{c}^{+}$ coming from the Cabibbo suppressed decay rate.
When applied to beauty decays, Voloshin's approach leads to a 
difference of $14\%$ in the lifetime of $\Xi_{b}^{-}$ with respect to the lifetime of 
$\Lambda_{b}$.

In this paper we extend Voloshin's analysis introducing a rather
moderate model dependence, in order to obtain  more predictive power, e.g. the
values for the lifetimes of the $(\Lambda_{b},\Xi_{b})$ hyperon triplet and
the lifetime of $\Omega_{b}$. Basically, we express the decay rates in terms
of the nonrelativistic (NR) wave function at the origin $\Psi(0)$, the value
of which we determine using Voloshin's method.

Our starting point is the expression (\ref{eq:master}). 
It is argued \cite{Ural} that the beauty mass
which enters the expression (\ref{eq:master}) is a running mass $m_{b}(\mu)$. 
In the limit $\mu \rightarrow 0$, one obtains the {\em pole} mass which
is very often used in calculations. It would be perfectly
legitimate to use the pole mass in pure perturbative theory (with no
nonperturbative contribution). 
%\begin{equation}
%\label{eq:pert}
%\Gamma_{SL}^{pert} =
%(m_{b}^{pole})^{5}[1-\frac{2\alpha_{s}(m_{b}^{2})}{3\pi}(\pi^{2}-
%\frac{25}{4})+...] \, .
%\end{equation}
% On the other
%side the $\alpha_{s}$ correction in (\ref{eq:pert}) is also the running
%$\alpha_{s}$ at some different scale $\nu$ -- this correction,
%$\alpha_{s}(\nu)$, is generated at high momenta, $\nu \gg \mu$. The scale
%$\mu$ is basically a factorization scale which separates short- and
%long-distance dynamics in the OPE. 
However, the use of the pole mass is
very problematic when nonperturbative corrections are calculated, because of
the renormalon singularities resulting in an irreducibile uncertainty of
$O(\Lambda_{QCD}/m_{b})$ that is larger than the nonperturbative corrections we are
calculating. Shielding $m_{b}(\mu)$ against renormalon ambiguities by
choosing $\mu > 0.5 \, GeV$, one avoids problems with the pole mass. In
fact,
a natural choice for the scale $\mu$ is $m_{b}/5 \sim 1 \, GeV$, as argued
in \cite{Big}. Such a relatively low scale makes the $\overline{MS}$ mass
inadequate for treating the decays. A natural definition of the running mass
would be that with the linear dependence on $\mu$:
\begin{equation}
\label{eq:run}
\frac{d \, m(\mu)}{d \, \mu} = -c_{m} \frac{\alpha_{s}(\mu)}{\pi} + ... \, .
\end{equation}

The recent value for $m_{b}(\mu)$ at $\mu \sim 1 \, GeV$ and for $c_{m}=16/9$
is given by \cite{BenSig,Big2} 
\begin{equation}
\label{eq:runmass}
m_{b}(\mu=1\, GeV) = (4.59 \pm 0.08) \, GeV \, ,
\end{equation}

which is slightly lower than the usual values. In this calculation we 
use $m_{b}(\mu=1\, GeV)$ in the range $4.6 \, GeV< m_{b}(1\, GeV) < 4.8 \,
GeV$.

Next, we turn to the calculation of the matrix elements of the $O_{6}^{i}$ 
(four-quark) operators. We follow the approach given by Voloshin
\cite{VolSU3}
based on HQET and flavor $SU(3)$ symmetry. 
A suitable parameter to express these matrix elements
is {\em the effective decay constant}  $F_{B}^{eff}$ which is an analogue
of the static decay constant used to evaluate four-quark matrix
elements in decays of heavy baryons \cite{BlokShif, GM, GMS}.

We use the following two differences of decay
rates:
For $\Delta_{b}^{1}=\Gamma(\Lambda_{b})-\Gamma(\Xi_{b}^{0})$, we have
\begin{equation}
\label{eq:D1}
\Delta_{b}^{1}=\frac{G_{F}^{2} m_{b}^{2}}{4 \pi} \mid V_{cb} \mid^{2}
(c^{2}-s^{2})(\sqrt{1-4z}-(1-z)^{2}(1+z))[C_{5}(m_{b}) x +C_{6}(m_{b}) y]
\/,
\end{equation}
where $z = m_{c}^{2}/m_{b}^{2}$ and $c$ and $s$ stand for $\cos \theta_{c}$
and $\sin \theta_{c}$, respectively ($\theta_{c}$ is the Cabibbo angle).
For $\Delta_{b}^{2}=\Gamma(\Xi_{b}^{-})-\Gamma(\Lambda_{b})$, we
have
\begin{equation}
\label{eq:D2}
\Delta_{b}^{2}=\frac{G_{F}^{2} m_{b}^{2}}{4 \pi} \mid V_{cb} \mid^{2}[l_{1}
x + l_{2} y] \/,
\end{equation}
where $l_{1}$ and $l_{2}$ are the abbreviations for the following
expressions:
\begin{equation}
\label{eq:l1}
l_{1}=(1-z)^{2}C_{1}(m_{b})-[c^{2}\sqrt{1-4z}+s^{2}(1-z)^{2}(1+z)]C_{5}(m_{b})
\/,
\end{equation}
\begin{equation}
\label{eq:l2}
l_{2}=(1-z)^{2}C_{2}(m_{b})-[c^{2}\sqrt{1-4z}+s^{2}(1-z)^{2}(1+z)]C_{6}(m_{b})
\/.
\end{equation}
In the equations displayed above, $C_{i}$ stand for special combinations of
Wilson
coefficients described in \cite{VolSU3}, while $x$ and $y$ denote 
combinations of heavy-baryon matrix elements introduced first in 
the same reference:
\begin{equation}
\label{eq:x}
x = \Big\langle \frac{1}{2} (\overline{b} \Gamma^{\mu} b)[(\overline{u}
\Gamma_{\mu} u) - (\overline{s} \Gamma_{\mu} s)] \Big\rangle
_{\Xi_{b}^{-}-\Lambda_{b}} =  \Big\langle \frac{1}{2} (\overline{b} \Gamma^{\mu}
b)[(\overline{s}
\Gamma_{\mu} s) - (\overline{d} \Gamma_{\mu} d)]
\Big\rangle_{\Lambda_{b}-\Xi_{b}^{0}} \, ,
\end{equation}  
\begin{equation}
\label{eq:y}
y = \Big\langle \frac{1}{2} (\overline{b^{i}} \Gamma^{\mu}
b^{j})[(\overline{u^{j}}
\Gamma_{\mu} u^{i}) - (\overline{s^{j}} \Gamma_{\mu} s^{i})] \Big\rangle
_{\Xi_{b}^{-}-\Lambda_{b}} =  \Big\langle \frac{1}{2} (\overline{b^{i}}
\Gamma^{\mu}
b^{j})[(\overline{s^{j}}
\Gamma_{\mu} s^{i}) - (\overline{d^{j}} \Gamma_{\mu} d^{i})]
\Big\rangle_{\Lambda_{b}-\Xi_{b}^{0}} \, ,
\end{equation}  
Similar relations are valid (through HQET and $SU(3)$ symmetry) for the
respective members of the charmed hyperon triplet \cite{VolSU3}.

The procedure of extraction of the effective parameter $F_{B}^{eff}$
is based on
equating expressions obtained in two approaches. In the first approach, for
the matrix elements $x$ and $y$ we use the $SU(3)$ hypothesis which basically
comprises using values of matrix elements extracted from experimental
data on charmed baryons for calculations in the beauty-baryon sector. This 
approach is based on the assumptions of $SU(3)$ and heavy-quark symmetry.
 In the second approach, $x$ and $y$ are   
calculated using the nonrelativistic quark model, already frequently employed   
for similar calculations \cite{BlokShif, GM, GMS}. Within this
model, for  $x$ and $y$ we have
\begin{equation}
\label{eq:valen}
x = -y = -\mid\Psi_{\Lambda_{b}}(0)\mid^{2}
\end{equation}
Equation (\ref{eq:valen}) clearly shows that the valence approximation is
used in the 
calculation of the matrix elements. The connection between the wave function
squared, 
$\mid\Psi_{\Lambda_{b}}(0)\mid^{2}$, and $F_{B}^{eff}$ is given by the
relation \cite{Ru, Cor}
\begin{equation}
\label{eq:psi}
\mid\Psi_{\Lambda_{b}}(0)\mid^{2} = T (F_{B}^{eff})^{2} \/,
\end{equation}
where
\begin{equation}
\label{eq:T}
T = 4 \frac{M(\Sigma_{b}^{0})-M(\Lambda_{b}^{0})}{M^{2}(B^{*})-M^{2}(B)}
m_{u}^{*} (\frac{1}{12} M(B) \kappa(\mu)^{-\frac{4}{9}}) \/.
\end{equation}
Here $\mu \sim 1 \, GeV$ is a typical hadronic scale of hybrid
renormalization $\kappa$.
The decay rate differences obtained in the
first and in the second approach are denoted by $\Delta_{b,SU(3)}^{i}$ and
$\Delta_{b,model}^{i}$, respectively ($i=1,2$).

The effective parameter $F_{B,i}^{eff}$ is now extracted from the equation
\begin{equation}
\label{eq:delta}
\Delta_{b,SU(3)}^{i}=\Delta_{b,model}^{i} \/.
\end{equation}
The expressions obtained for $F_{Beff,i}$, $i=1,2$, are
\begin{equation}
\label{eq:extr1}
F_{B,1}^{eff} = \sqrt{\frac{C_{5}(m_{b}) x +C_{6}(m_{b})
y}{T (C_{6}(m_{b})-C_{5}(m_{b}))}} \/,
\end{equation}
\begin{equation}
\label{eq:extr2}
F_{B,2}^{eff} = \sqrt{\frac{l_{1} x +l_{2} y}
{T (l_{2}-l_{1})}} \/.
\end{equation}

The final numerical value is calculated taking into consideration the errors
of the expressions (\ref{eq:extr1}) and (\ref{eq:extr2}) and combining all
numerical values appropriately.
For $m_{c}=1.25 \, GeV$ and $m_{b}=4.7 \, GeV$, we
obtain
\begin{equation}
\label{eq:FB}
F_{B}^{eff} = (0.441 \pm 0.026) \, GeV \, .
\end{equation}
The parameter $F_{B}^{eff}$ shows a slight mass dependence which was incorporated in
numerical calculations.

\begin{figure}[h]
\centerline{\resizebox{0.6\textwidth}{!}{\includegraphics{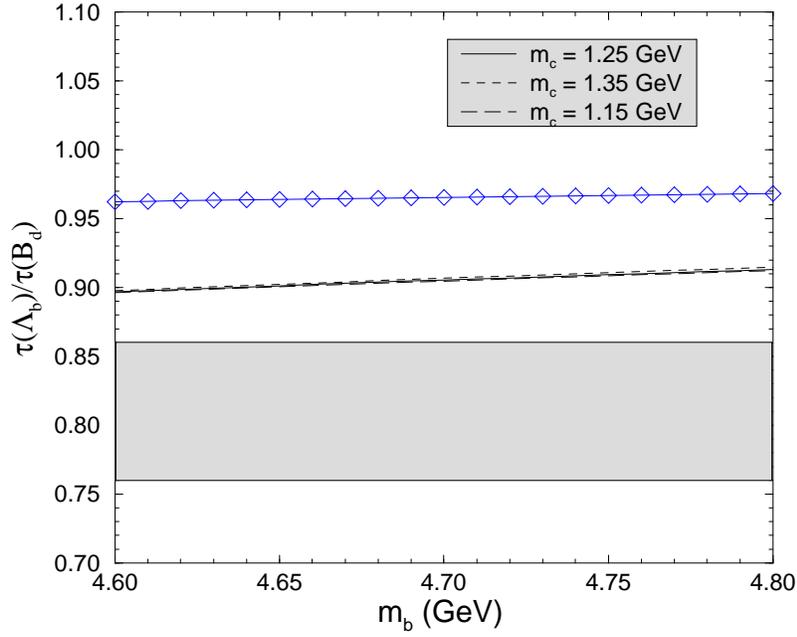}}}
\caption{\label{fig:ratio}
\normalsize The shaded area represents the experimental value of
the ratio $r_{\Lambda B}$ within one standard deviation.
The line with diamonds represents the calculated value of
$r_{\Lambda B}$ for the "standard" value $f_{B} = 160 \, MeV$.
The values of $r_{\Lambda B}$ using $F_{B}^{eff}$ are calculated for three
different values of mass $m_{c}$ and are represented by lines without
symbols. The significant shift from the "standard" value  result is
visible, but the deviation from the experimental band is still substantial.}
\end{figure}

Next, the numerical value displayed in (\ref{eq:FB}) is used in (\ref{eq:psi})
to obtain the values of the $O_{6}^{i}$ matrix elements. As all the matrix
elements in the expression (\ref{eq:master}) for the decay rate are now available, we
can calculate the lifetimes of beauty baryons accordingly. The lifetimes of
$B$ mesons are calculated in a "standard" way using the B-meson decay constant $f_{B}$.

Since absolute results for lifetimes are not so reliable owing to ambiguities
in quark mass, we shall express our results mainly in the form of lifetime 
ratios. 

Our results for the ratio $r_{\Lambda B} =
\tau(\Lambda_{b})/\tau(B_{d}^{0})$ are shown in Fig.\ref{fig:ratio}. 
The Wilson coefficients in (\ref{eq:master}) have been calculated at one loop
using $\Lambda_{QCD} = 300 \, MeV$. Other relevant numerical parameters used
throughout the paper are \cite{Neu, BlokShif} $\mu_{\pi}^{2} = 0.5 \,
GeV^{2}$,
$\mu_{G}^{2}(\Omega_{b}) = 0.156 \, GeV^{2}$, $\mu_{G}^{2}(\Lambda_{b}, \Xi_{b}) = 0$.
The effect of introducing $F_{B}^{eff}$, which we have calculated, is to bring the ratio
$r_{\Lambda B}$ from $0.96$ to $0.9$ -- still two standard
deviations from experiment. It is clear from Fig.\ref{fig:ratio} that
variation in $m_{b}$ has almost no effect. Also, the variation of
$\mu_{\pi}^{2}$ does change $r_{\Lambda B}$ at the permile level. If the
experimental ratio (\ref{eq:ratioexp2}) persists, then
there might be the problem in $b$ decays. 

\begin{figure}[h]
\centerline{\resizebox{0.6\textwidth}{!}{\includegraphics{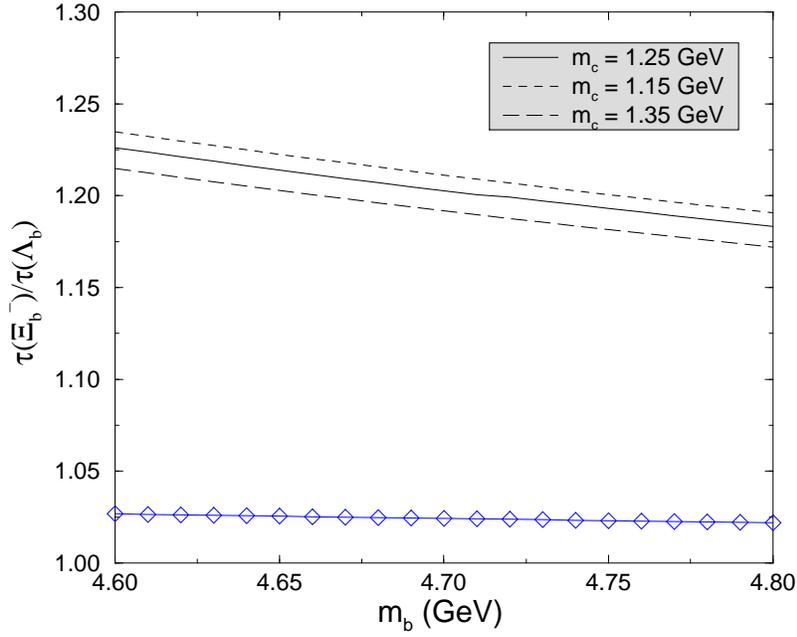}}}
\caption{\label{fig:ksi} The line with diamonds represents the value of
the ratio $\tau(\Xi_{b}^{-})/\tau(\Lambda_{b})$ for the value $f_{B} = 160 \,
MeV$. Values of
the same ratio calculated using $F_{B}^{eff}$ are given for three $m_{c}$
masses and are represented by lines without symbols. The difference of
$\sim 20 \%$ is clearly visible.}
\end{figure}

An interesting effect in this approach noticed by Voloshin is the enhancement
of the lifetime of $\Xi_{b}^{-}$ compared with the lifetime of $\Lambda_{b}$.
Using the "standard" $B$-meson decay constant $f_{B} \sim 160 \, MeV$, instead of
$F_{B}^{eff}$, one obtains for the ratio
$\tau(\Xi_{b}^{-})/\tau(\Lambda_{b}) \sim 1.03$. Our calculation gives, 
Fig.\ref{fig:ksi},
\begin{equation}
\label{eq:ksilam}
\frac{\tau(\Xi_{b}^{-})}{\tau(\Lambda_{b})} \simeq 1.22 \, ,
\end{equation}
i.e. a relative enhancement of the
$\tau(\Xi_{b}^{-})/\tau(\Lambda_{b})$ ratio of the order  $20 \%$,
which is in fair agreement with the preliminary experimental results
(\ref{eq:totals}) \cite{PDG}.
The main reason for this enhancement is the large (positive) exchange 
contribution to $\Gamma(\Lambda_{b})$ versus the large negative Pauli 
interference contribution in $\Gamma(\Xi_{b}^{-})$.
 
\begin{figure}
\centerline{\resizebox{0.6\textwidth}{!}{\includegraphics{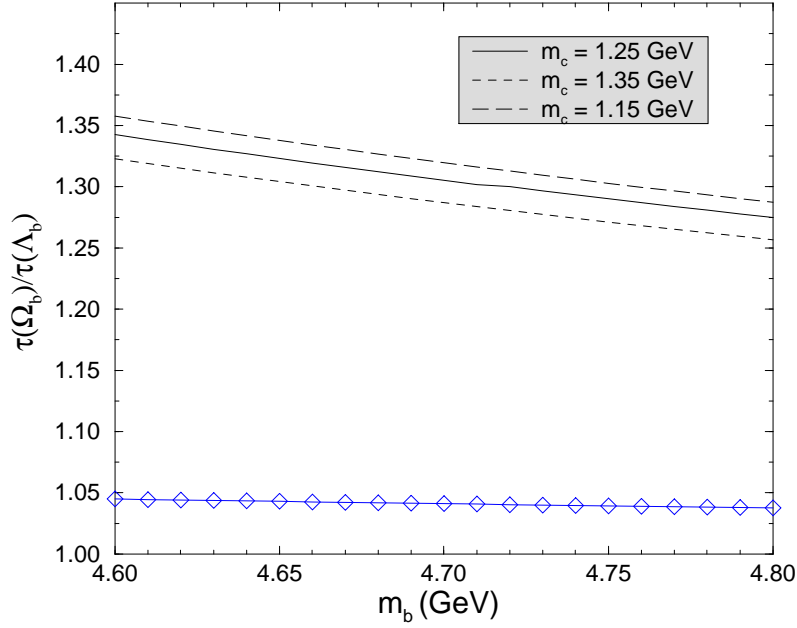}}}
\caption{\label{fig:omega} The value of the ratio
$\tau(\Omega_{b})/\tau(\Lambda_{b})$ obtained using $f_{B} = 160 \, MeV$
is represented
by the line with diamonds. Values of the same ratio for $F_{B}^{eff}$,
represented by lines without symbols, are calculated for three different
$m_{c}$ masses. Calculations using $f_{B}$ and $F_{B}^{eff}$ differ by $\sim 30
\%$.}
\end{figure}

Such an enhancement is even more pronounced in the lifetime of $\Omega_{b}$,
giving the ratio
\begin{equation}
\label{eq:omlam}
\frac{\tau(\Omega_{b})}{\tau(\Lambda_{b})} \simeq 1.35 \, ,
\end{equation}
i.e. a relative enhancement of the ratio $\tau(\Omega_{b})/\tau(\Lambda_{b})$ 
of the order $30 \%$, Fig.\ref{fig:omega}.
This result is a consequence of the even stronger (compared with $\Xi_{b}^{-}$)
negative Pauli interference contribution in $\Gamma(\Omega_{b})$. 
The results (\ref{eq:ksilam}) and (\ref{eq:omlam}) should both serve as 
crucial predictions to be checked in  experiment.
The calculation gives $\tau(\Xi_{b}^{0})$ approximately the same as
$\tau(\Lambda_{b})$, which together with (\ref{eq:ksilam}) and 
(\ref{eq:omlam}) leads to the following lifetime hierarchy:
\begin{equation}
\label{eq:hierarchy}
\tau(\Lambda_{b}) \simeq \tau(\Xi_{b}^{0}) < \tau(\Xi_{b}^{-}) <
\tau(\Omega_{b}) \, .
\end{equation}
As long as the absolute value of the $\Lambda_{b}$ lifetime is concerned, the
effect of $F_{B}^{eff}$ is to lower the theoretical value of 
$\tau(\Lambda_{b})$ by $\sim 10 \%$, giving
\begin{equation}
\label{eq:lambda}
\tau(\Lambda_{b}) \sim 2.0 \, ps \, ,
\end{equation}
which is too high compared with the measured value $ (1.24 \pm 0.08) \, ps$. To
obtain better agreement with experiment, one needs larger $m_{b}$. If, 
for example, we use
the pole masses $m_{b}^{pole} = 5.1 \, GeV$, $m_{c}^{pole} =
1.6 \, GeV$, the result is $\tau(\Lambda_{b})^{pole} = 1.6 \, ps$ -- still
too high vis-\`{a}-vis experiment. However, playing with a large pole mass is
merely an introduction of an additional parameter -- a consistent treatment
requires having the running mass $m_{b}(\mu)$ in the expansion and its value
could hardly reach more than $4.7 \, GeV$.

Much the same situation appears in the calculation of B-meson lifetimes,
which is not affected by our approach. Typically, one obtains $\tau(B) \sim
2-2.5 \, ps$ for $m_{b} = 4.6 \, GeV$ and $\tau(B) \sim 1.75-2.2 \, ps$ for
$m_{b} = 4.7 \, GeV$, the range of values for $\tau$s coming from the
variation of $m_{c}$, $1.15 \, GeV< m_{c} < 1.35 \, GeV$. 
Comparing with the results 
for the calculated value of $\tau(\Lambda_{b})$, one sees that it is
easier to have $\tau(B_{d}^{0})$ near to the experimental value. This may
suggest that the problem with too large a theoretical value of 
$r_{\Lambda B}$ lies
in the theoretical overestimate of $\tau(\Lambda_{b})$.

To conclude, we point out the following. 
The calculations presented in this
paper rely upon HQET and flavor $SU(3)$ symmetry and are therefore reliable
up to violations of these symmetries. Still, we expect the effects of these 
violations to be smaller than the main effect of our approach.
The procedure applied above significantly increases
the contribution of four-quark operators and 
numerical results show a significant, albeit still unsufficient shift 
towards experimental values, especially in the case of the $r_{\Lambda B}$
ratio.
%Although we achieved that the baryonic wave function squared 
%$|\Psi_{\Lambda_b}(0)|^2$ is more then four times bigger than its mesonic 
%counterpart, $|\Psi_B(0)|^2$, we are still far from the required ratio
%$|\Psi_{\Lambda_b}(0)|^2/|\Psi_B(0)|^2 \sim 16$ to approach the 
%experimental findings from (\ref{eq:ratioexp2}). 
In our approach, to reach the experimental value of $r_{\Lambda B}$,
 would require $F_{B}^{eff}$ to 
have the value $0.72 \, GeV$, which can be hardly achieved. 
The discrepancy, still remaining after increasing the preasymptotic effects 
coming
from four-quark operators, indicates that there should be other, yet
unknown, sources of enhancement of preasymptotic effects and that these
effects should also produce significant contributions. Also, there remains the
possibility of violation of some of the underlying concepts, such as
quark-hadron duality, but a consistent treatment of these problems is still
out of the reach of the present theory.
In our approach,
the large contributions of the operators of $D=6$ also suggest 
a much wider spread of lifetimes in the sector of beauty baryons. The extent
of this spread is to be verified by future experiments.
At the end, we state that a systematic application of the OPE, HQET and
a moderately model-dependent procedure of enhancement of preasymptotic effects
improves the $r_{\Lambda B}$ ratio significantly,
although it cannot resolve the problem completely. We consider this
problem along with the problem of absolute lifetimes of beauty hadrons to be
one of the most important issues that heavy-quark physics should address
in the future.     

Acknowledgements 

This work was supported by the Ministry of
Science and Technology of the Republic of Croatia under the contract Nr.
00980102.

\end{document}